\documentclass[
    twocolumn,
    10pt,
    a4paper,
    float,
    aps,
    pra,
    nofootinbib
]{revtex4-2}

\usepackage{amsmath}
\usepackage{amssymb}

\usepackage[utf8]{inputenc}

\usepackage[dvipsnames]{xcolor}
\usepackage{hyperref}
\hypersetup{
    colorlinks  =   true,
    citecolor   =   BrickRed,
    linkcolor   =   RoyalBlue
}

\usepackage{float}
\usepackage{graphicx}
\usepackage{dcolumn}
\usepackage{cancel}
\usepackage{ulem}

\usepackage{bm}
\usepackage{bbm}
\usepackage{bbold}
\DeclareMathAlphabet{\mathbbb}{U}{bbold}{m}{n}

\usepackage{mathrsfs}

\newcommand{\eff}{\text{eff}}

\newcommand{\la}{\langle}
\newcommand{\ra}{\rangle}
\newcommand{\tp}{\mathsf{T}}
\newcommand{\nex}{n_{\text{ex}}}

\newcommand{\customsec}[1]{
     \textbf{\textit{ #1}}
}

\DeclareMathOperator{\haf}{haf}

\DeclareMathOperator{\Tr}{Tr}

\begin{document}

\title{\large Boson sampling with self-generation of squeezing via interaction of photons and atoms}

\author{S.V. Tarasov}
\affiliation{Institute of Applied Physics, Russian Academy of Sciences, Nizhny Novgorod 603950, Russia}
\author{Vl.V. Kocharovsky}
\affiliation
{Institute of Applied Physics, Russian Academy of Sciences, Nizhny Novgorod 603950, Russia}

\date{\today}
\begin{abstract}
We suggest a novel scheme for generating multimode squeezed states 
for the boson sampling implementation. 
The idea is to replace a commonly used linear interferometer by a multimode resonator containing a passive optical element consisting of two-level atoms dispersively interacting with photons and self-generating a squeezed compound state of both bosons – photons and atoms. 
The suggested scheme does not need (a) on-demand external sources of photons in squeezed or Fock quantum states and (b) numerous interchannel couplers which introduce phase noise and losses that prevent scaling up the system and achieving quantum advantage. 
The idea is illustrated by a setup based on a Bose-Einstein-condensed gas confined in a multimode resonator, one of whose optical modes is in the classical coherent regime.
The joint probability distribution of photon and/or noncondensed atom numbers is calculated via a matrix hafnian that, for certain parameters of the system, is hardly to be effectively calculated by classical computers. 
Such experiments are at reach via existing cavity-QED and cold-gas technology.
      
\end{abstract}
\maketitle	
	

\customsec{Challenging quantum advantage with boson sampling,}
based on a quantum interference of indistinguishable boson particles, has been proposed and is widely discussed nowadays in the race to demonstrate advantage of quantum simulators over classical computers~\cite{LundRalphNPj2017,MontanaroNat2017}. 
It is motivated by the fact that, for multimode interferometers whose input modes are fed with proper photon states, the joint probability distribution of bosons over output modes is related to functions which, in general case, are unlikely to be effectively calculated or approximated by classical computers.
Examples of such functions are matrix permanents~\cite{Aaronson2011,Valiant1979-PerComplexity,Barvinok-Combinatorics}, hafnians~\cite{Hamilton2017-GBS,Barvinok-Combinatorics}, loop hafnians~\cite{Bjorklund2019}, torontonians~\cite{Quesada2018-GBSthreshold}, etc.
It is supposed that for the classical computation of all these functions the number of operations grows exponentially with the size of the matrix in the argument.

The concept of boson sampling has a straightforward implementation in the quantum optics framework.
However, demonstrating the quantum advantage is challenging task since the system should demonstrate proper scalability. 
The number of modes and the number of photons exiting the interferometer both should be sufficiently large, so that the matrices associated with outcome probabilities turn out to be of a large dimension, such that their permanents and hafnians are inaccessible to classical computation.
Also, the interferometer should be deep enough to effectively mix all the input states and redistribute them among all of the output channels.

Among the main factors which limit scalability are photon losses.
Obviously, they may prevent one from achieving big enough number of photons exiting the setup.
Exactly this has happened with the experiments on the original boson sampling with single-photon input states suggested in \cite{Aaronson2011}.
The best attempt has a $60$-mode interferometer, up to $20$ input photons and $14$ detected photons \cite{Wang2019-expBS}.
These numbers are not big enough, since the outcome probabilities are related to permanents of matrices whose dimension is equal to the total number of photons on interferometer outputs, and permanents of up to $60 \times 60$ arbitrary matrices are already accessible for a straightforward calculation via classical super-computers \cite{Wu2016-TianhePer}.
That restriction may be overcome for modified protocols. 
One of the most advanced variations is the so-called Gaussian boson sampling, which operates with squeezed vacuum photons simultaneously fed to the input modes of interferometer and number-resolving detectors counting photons in the output channels \cite{Hamilton2017-GBS,Hamilton2019-DetailedGBS}.
The use of squeezed light instead of single-photon Fock input states helped to implement experiments of larger scales, with up to 144 mixed modes and 113 detected photons in average \cite{Pan2021-GBSexperiment113}, or up to 216 mixed modes and 125 detected photons in average \cite{QuesadaLavoie2022-GBS}.
However, the loss of photons leads to another restriction.

Beyond the aforementioned difficulty, there is even more important one.
The presence of the high photon loss rate opens up several ways to efficiently mimic the boson sampling via classical computing devices.
Indeed, for a original boson sampling, if the number of photons survived passing through interferometer scales slower than the square root of number of input photons, an approximation via thermal states (for deep interferometers) and quasi-polynomial tensor-product calculations (for shallow interferometes) work well \cite{Shchesnovich2019-lossyBS}; an approximation involving separable states also applies \cite{Oszmaniec2018-lossyBS,Oszmaniec2020-lossyBS}.
With the same boundary of the loss rate, the Gaussian boson sampling also allows for an effective classical simulation. The relevant tensor-network approach has been demonstrated in \cite{Alexeev2023,Alexeev2024}.  

Since the deep interferometers are based on lossy beam splitters and suffer from exponential decay of the transmission with the circuit depth, a possibility to implement quantum advantage in boson sampling experiments via typical setups looks questionable \cite{Shchesnovich2019-lossyBS}. So, the novel, beyond the aforementioned circuit, architectures bypassing the above limitations are needed.

The present paper is dedicated to a novel system for boson sampling, which doesn't require numerous interchannel couplers to produce multimode squeezed and entangled photon states.
Also, it doesn't involve a large number of on-demand external sources of photons, and therefore avoids the other serious problems of current boson sampling schemes related to synchronizing and phase locking the input photon sources.

Recently we found that the Gaussian boson sampling process is essentially incarnated by multimode atom number fluctuations of an equilibrium, partially condensed weakly-interacting Bose gas \cite{PRA2022-AtomicBS}. 
Namely, for the joint statistics of occupation numbers of a set of different excited (noncondensate) single-particle states, the corresponding joint probabilities of atom numbers are proportional to hafnians of certain matrices, similar to the case of an optical Gaussian boson sampling setup.
The non-trivial hafnians arise even in the case of thermal equilibrium and even within the mean-field description due to presence in the Hamiltonian of the so-called counter-rotating terms (which are products of two annihilation or two creation operators) along with the co-rotating terms (which are products of creation and annihilation operators).
The point is that these counter-rotating terms enable squeezing effects in the transition between quasiparticles of the system, which fluctuates independently, and atoms, which occupation numbers $n_j$ in different, $j$-th modes are being sampled.

That has been shown via an explicit calculation of the characteristic function of the joint atom numbers statistics in a gas of atoms described by a two-body interaction Hamiltonian $\hat{\mathcal{H}}$ and an equilibrium Gaussian state with the density matrix $\hat{\rho} \propto \exp(-\hat{\mathcal{H}}/T)$ at temperature $T$:
\begin{equation}
\begin{split}
    &\Theta\big(\{z_j\}\big) \equiv \Tr \Big(\hat{\rho} \prod_j z_j^{\hat{n}_j} \Big) = \frac{1}{\sqrt{\mathbb{1} + G}} \frac{1}{\sqrt{\mathbb{1} - Z G(\mathbb{1}+G)^{-1}}},
    \\
    &\qquad Z   \equiv  \begin{bmatrix}
                            \textrm{diag}\big(\{z_j\}\big) & \mathbb{0} \\
                            \mathbb{0} & \textrm{diag}\big(\{z_j\}\big)
                        \end{bmatrix},
    \qquad
    G = \begin{bmatrix}
            \mathcal{N} & \mathcal{A}^* \\
            \mathcal{A} & \mathcal{N}^*    
\end{bmatrix}.
\end{split}    
\end{equation}
Symbols $\mathbb{1}$ and $\mathbb{0}$ hereinafter stand for identity and zero matrices of a suitable dimension, respectively.
The characteristic function of a Gaussian state is fully determined by the covariance matrix $G$ which consists of all pair correlations of creation/annihilation operators describing the modes under investigation. Its blocks $\mathcal{N} = \big( \la \hat{b}_j^\dagger \hat{b}_k \ra \big)$ and $\mathcal{A} = \big( \la \hat{b}_j \hat{b}_k \ra \big)$ describe normal and anomalous correlations of creation and annihilation operators $\hat{b}_j^\dagger$ and $\hat{b}_j$. The indices $j$ and $k$  enumerate all excited atomic modes included in the consideration.
The characteristic function $\Theta\big(\{z_j\}\big)$ coincides -- up to a scaling factor -- with a hafnian generating function, in accord with the hafnian master theorem we recently found \cite{LAA2022-HafnianMasterTheorem}.
Hence, the joint probabilities of the mode occupation numbers, which are by definition determined by the mixed derivatives of the characteristic function,
\begin{equation}
        p(\{n_j\}) = \prod_j \frac{\partial^{n_j}}{n_j! \, \partial z_j^{n_j}} \, \Big. \Theta\big(\{z_j\}\big) \Big|_{\{z_j=0\}},
\end{equation}
are proportional to hafnians of certain covariance-related matrices, as in the case of the Gaussian boson sampling.
It is important to emphasize the role of counter-rotating terms in the Hamiltonian $\hat{\mathcal{H}}$, since these terms are responsible for squeezing effects and appearance of nonzero anomalous correlations, $\mathcal{A} \neq \mathbb{0}$.
With a Hamiltonian consisting of co-rotating terms only, sampling probabilities are not believed being truly hard to simulate: 
In the case of vanishing anomalous correlators, the matrix hafnians would be reduced to permanents of positive-definite matrices, which may by approximated via the Stockmeyer algorithm \cite{Stockmeyer1985-ApproximationForNP} as has been stated in \cite{Ralph2015PRL}.

In the present paper, we propose to generate multimode squeezed photonic states suitable for Gaussian boson sampling right in the multimode cavity via employing the same mechanism as for the atomic boson sampling, instead of generating squeezed states by an external sources, synchronizing their input and mixing them via interferometer.
The idea is to fill the cavity supporting a number of high-Q modes with a medium introducing counter-rotating terms in the system's description.
As an example of such a medium, we consider a cold, partially condensed gas of two-level Bose atoms interacting with photons in the non-resonant way.
The counter-rotating terms associated with the light-atom interaction bring the desired squeezing effects to the hybrid modes of the system, and the coupling existing between atoms and light distributes that squeezing to photons.

The paper is organized as follows.
First, we recall the standard description of a multimode cavity-QED hybrid atom-photon system, and introduce a Gaussian ansatz describing its pseudo-equilibrium state.
Further, we discuss the corresponding joint multimode quantum statistics in the photon subsystem: The probabilities of photon occupation numbers are proportional to matrix hafnians which, in general, are hard-to-compute in a wide range of parameters of the system providing large enough values of anomalous correlators.
Finally, employing a simple model of two modes -- one atomic and one optical -- as an example, we demonstrate that such large anomalous correlators are indeed available.

\customsec{Description of a multimode cavity-QED system}
should be started with stating its main feature: The cavity supports a number of high-Q photon modes enumerated by an integer index $\nu$. These modes are characterized by frequencies $\omega_\nu$ and profiles of the electric field $\bm{\mathcal{E}}_\nu({\bf r})$, normalized to a single photon energy according to $\int |\bm{\mathcal{E}}_\nu({\bf r})|^2 d^3{\bf r} = 2\pi \hbar \omega_\nu$.
There is a single mode, denoted by $\nu = 0$, which is macroscopically populated, so that the corresponding electric field ${\bf E}_0({\bf r})$ of the frequency $\omega _0$ describes a classical coherent light.
This mode could be either a driving light wave, typically generated by an external source in cavity-QED experiments \cite{cavityQED-Humb,cavityQED-Stanford,cavityQED-ETH1,cavityQED-ETH2}, or a kind of photon condensate similar to that established in experiments with dye-filled resonators \cite{Klaers2010lightBEC,Wang2019lightBEC}.
Other modes, $\nu \ge 1$, are populated on a low, quantum level.
The photon ladder operators adding or removing one photon to or from the $\nu$-th mode are denoted by $\hat{a}^\dagger_\nu$ and $\hat{a}_\nu$, respectively, and the operator of the number of photons in the mode is $\hat{a}^\dagger_\nu \hat{a}_\nu$.
For further convenience, we also define a number of photons in a coherent light mode, $Q_0 = \big( \int |{\bf E}_0({\bf r})|^2 d^3{\bf r} \big) \big/ 2\pi\hbar\omega_0 \gg 1$, which is the first macroscopically large parameter regarding the system.

We assume that cold, partially condensed Bose gas is trapped inside the cavity by an external potential $V_{tr}({\bf r})$, so that the atomic cloud is well-overlapped with the high-Q modes.
Two-level atoms, whose internal structure is represented by the lower energy state $| g \ra$ and the upper energy state $| e \ra$, have a transition frequency denoted by $\omega_a$, the dipole moment ${\bf d}$, and the decay rate $\tau_2^{-1}$.
Basically, ensemble of two-level atoms is described via two field operators $\hat\psi_g({\bf r})$ and $\hat\psi_e({\bf r})$, one per lower and upper intrinsic states, respectively.
We make a standard assumption that all optical frequencies $\omega_\nu$ are well-detuned from the resonance meaning that the frequency differences $\Delta_a \equiv \omega_a-\omega_0$ and $\Delta_\nu \equiv \omega_0 - \omega_\nu$ obey the following inequalities:
$|\Delta_a|, |\Delta_a + \Delta_\nu| \gg \tau_2^{-1}$.
That assumption allows one to neglect by effects of spontaneous emission and describe the system in terms of the lower-energy-state field operator $\hat\psi_g({\bf r}) \equiv \hat\psi({\bf r})$ only, while the upper-energy-state field operator $\hat\psi_e({\bf r})$ may be adiabatically excluded \cite{Ritsch2008-AtomsInLattices}.

The gas is partially condensed, thus the field operator can be effectively decomposed according to the Bogoliubov approximation: $\hat{\psi}({\bf r}) \simeq \phi_0({\bf r}) \sqrt{N_0} + \sum_j f_j({\bf r}) \hat{A}_j$. 
Here $\phi_0({\bf r})$ is a single-particle wave function of a macroscopically populated condensate state with the mean occupation $N_0 \gg 1$.
The sum represents the noncondensed modes, or excited modes regarding the translation motion of atoms: The ladder operators $\hat{A}_j$ and $\hat{A}^\dagger_j$ annihilate and create an atom in the state with the wave function $f_j({\bf r})$, and all these functions are orthogonal to the condensate wave function $\phi_0({\bf r})$.
The functions $f_j({\bf r})$ and $\phi_0({\bf r})$ together constitutes a full basis in the Hilbert space of the single-particle wave functions. 

\customsec{The Hamiltonian of the interacting atoms and photons}
in a frame rotating with the frequency $\omega_0$ of the coherent light field has the following standard form~\cite{Ritsch2021-CavityQED}:
\begin{widetext}
    \begin{equation}    \label{H-cavityQED}
    \begin{split}
        \hat{\mathcal{H}} &= \int \hat{\psi}_a^\dagger \left( -\frac{\hbar}{2m}\nabla^2 + V_{tr}(r) + \frac{\hbar |\Omega(r)|^2}{\Delta_a} + \frac{g_a}{2} \hat{\psi}_a^\dagger \hat{\psi}_a \right) \hat{\psi}_a \, d^3 {\bf r}
        \\
        &+\int \hat{\psi}_a^\dagger\frac{\hbar}{\Delta_a} \sum_\nu
            \big( \mathcal{G}_\nu^* \, \Omega \, \hat{a}_\nu^\dagger
                    + \mathcal{G}_\nu \, \Omega^* \, \hat{a}_\nu \big) \hat{\psi}_a \, d^3{\bf r}
        +\int \hat{\psi}_a^\dagger\frac{\hbar}{\Delta_a} \sum_{\nu,\nu'}
            \big( \mathcal{G}_\nu^* \, \mathcal{G}_{\nu'} \, 
                    \hat{a}_\nu^\dagger \, \hat{a}_{\nu'} \big)
                     \hat{\psi}_a \, d^3{\bf r} 
        -\sum_\nu \hbar \Delta_\nu \hat{a}_\nu^\dagger \hat{a}_\nu.
    \end{split}
    \end{equation}
\end{widetext}
Here the constant $g_a$ characterises the s-wave atom-atom scattering in a weakly interacting gas, and the interaction between atoms and photonic modes is characterized by Rabi frequencies: $\Omega(\bf{r}) = {\bf d}\, {\bf E}_0 /\hbar$ regarding the classical light mode ($\nu = 0$), and $\mathcal{G}_\nu(\bf{r}) = {\bf d}\, \bm{\mathcal{E}}_\nu /\hbar$ regarding the high-Q modes ($\nu \neq 0$) which are populated at a low, quantum level. 

The above cavity-QED system in fact incorporate two macrosopically populated modes, one atomic and one optical.
Thus, for the Hamiltonian~(\ref{H-cavityQED}) written in terms of the aforementioned basis of atomic and photon modes, all terms may be ranked over the order of magnitude according to two large parameters $Q_0 \gg 1$ and $N_0 \gg 1$.

The first summand in the Hamiltonian $\hat{\mathcal{H}}$ describes the weakly interacting gas in the confining potential $V_{tr}$ supplemented with an optical lattice potential created by the field of the classical coherent wave.
The second summand describes the interaction of atoms with the classical coherent optical mode. 
In the leading order (which involves two condensate wave functions and scales as $N_0\sqrt{Q_0}$) it corresponds to the process of photon scattering from the macroscopically populated mode into other, quantum-level populated high-Q ($\nu \neq 0$) optical modes of the cavity and the backward process. 
In the next-to-leading order, proportional to $\sqrt{N_0Q_0}$, it describes an effective interaction between different high-Q cavity modes and different spatial atomic states. 
Finally, the third summand stands for photon exchange between the $\nu$-th and $\nu'$-th modes due to scattering on atoms.

We include in the Hamiltonian the high-Q photon modes only. 
Their interaction with atoms, even in the case of a low photon number, is significant: It is strongly enhanced since the lifetime of a photon in the resonator significantly exceeds the time of traversing the resonator. (In other words, a photon has a large number of coherent acts of interaction with the atoms while passing through the cloud of atoms.)  
The low-Q modes experience no such amplification in interaction. Therefore, taking into account also their low intensity, they are omitted.

The behavior of the considered hybrid atom-photon system is determined by the Hamiltonian dynamics along with the slow rate dissipation, caused mainly by the photon leakage from high-Q modes including scattering of photons from the high-Q into the low-Q modes.
Typically, such an open system evolves towards some steady or quasi-steady state, which is characterized by nonzero mean photon numbers regarding all the high-Q modes due to scattering of a coherent light mode on the atomic cloud.
In general, that state does not coincides with a thermal one.
However, its properties in many cases may be approximated by a suitable thermal state with some effective temperature $T_\eff$, determined by dissipative parameters such as optical mode decay rates, noise intensity and so on.
Such an anzats can be justified for calculating distribution function for the population of the optical modes in a thermodynamic limit as is shown in \cite{Lebreuilly2018Pseudothermalization}.
Also, it looks applicable for calculating normal and anomalous correlators which characterize fluctuations in the optically-driven BEC within the Keldysh approach developed in \cite{Torre2013Keldysh}.
Remarkably, the pseudo-thermalization is predicted for non-Markovian reservoirs despite of their highly non-thermal nature and, in some cases, may not be restricted to low-frequency energies only.

Aiming to illustrate the idea of Boson sampling from a hybrid atomic-optical system in a simplest possible way, we assume that a quasi-steady state of the system is a Gaussian pseudo-thermalized state with an effective quadratic Hamiltonian whose structure is inherited from the Hamiltonian~(\ref{H-cavityQED}).
Namely, the condensate atomic wave function $\phi_0$ and the coherent optical mode ${\bf E}_0$ represent some self-consistent stationary solutions, guided by relevant Gross-Pitaevskii-type equations, and fluctuations of interacting atoms and photons in the excited atomic and high-Q optical modes, on top of those coherent fields, are described by the density matrix 
\begin{equation}    \label{rho-eff}
    \hat\rho \propto \exp(- \hat{\mathcal{H}}_\eff/T_\eff).
\end{equation}
Here the effective Hamiltonian $\hat{\mathcal{H}}_\eff$ involves the same terms regarding non-macroscopically populated modes (which are quadratic in atom ladder operators $\hat{A}_j$, $\hat{A}_j^\dagger$ and photon ladder operators $\hat{a}_\nu$, $\hat{a}_\nu^\dagger$) as the Hamiltonian~(\ref{H-cavityQED}):
\begin{equation}    \label{H-eff-2}
\begin{split}
    \hat{\mathcal{H}}_\eff 
        &= \frac{1}{2} 
            ({\bf a}, {\bf A}, {\bf a}^\dagger, {\bf A}^\dagger) 
                \, H \,
                    ({\bf a}^\dagger, {\bf A}^\dagger, {\bf a}, {\bf A})^\tp,
    \\
    H   &=  
    \renewcommand{\arraystretch}{1.5}
    \left[
    \begin{array}{cc|cc}
                \epsilon_{ph}^* + S_{ph}^*  &  S_{at-ph}^*    &   \mathbb{0}  &  \tilde{S}_{at-ph} \\
                S_{at-ph}^\tp &   \epsilon_{at}^* + S_{at}^*  &  \tilde{S}_{at-ph}^\tp    & \tilde{S}_{at} \\\hline
                \mathbb{0}   &  \tilde{S}_{at-ph}^*    &   \epsilon_{ph} + S_{ph}   & S_{at-ph} \\
                \tilde{S}_{at-ph}^\dagger  & \tilde{S}_{at}^* &   S_{at-ph}^\dagger    & \epsilon_{at} + S_{at}
    \end{array}
    \right].
    \renewcommand{\arraystretch}{1}
\end{split}
\end{equation}
We keep in consideration only a finite number of excited atomic modes $m_{at}$, assuming the occupation numbers in all of the omitted modes are small enough and could be neglected. 
The matrix $H$, which is sometimes called the grand-dynamical matrix \cite{Colpa1978QuadBosH}, is of an essentially $4\times 4$ block structure.
Here ${\bf A} \equiv (\hat{A}_1, \hat{A}_2, \hat{A}_3 \ldots)$ is a row vector which consists of atomic annihilation operators $\hat{A}_j$, while ${\bf a}$ is a row vector which consists of photonic annihilation operators $\hat{a}_{\nu}$.
${\bf A}^\dagger$ and ${\bf a}^\dagger$ are the similar row-vectors consisting of the creation operators.
In Eq.~(\ref{H-eff-2}), these operators are combined in an extended $2m_{at}+2m_{ph}$ vectors, including all creation and annihilation operators of either atomic or optical modes.
In the following analysis we omit all other terms which are of the higher than quadratic order in ladder operators of low-populated atomic and optical modes assuming that their contribution is relatively small.

Both co-rotating and counter-rotating atom-photon intermode $m_{ph}\times m_{at}$ blocks $S_{at-ph}$ and $\tilde{S}_{at-ph}$ originate from the second summand of the Hamiltonian~(\ref{H-cavityQED}), describing an effective interaction between different high-Q modes and different spatial atomic states:
\begin{equation}
\begin{split}
    &(S_{at-ph})_{\nu j} \, \hat{a}_\nu^\dagger \hat{A}_j 
    \equiv
    \frac{\hbar}{\Delta_a} \sqrt{N_0} \Big( \int \phi_0^* \Omega f_j \mathcal{G}_\nu^* \, d^3r\Big)  \, \hat{a}_\nu^\dagger \hat{A}_j,
    \\
    &(\tilde{S}_{at-ph})_{\nu j} \, \hat{a}_\nu \hat{A}_j
    \equiv
    \frac{\hbar}{\Delta_a} \sqrt{N_0} \Big( \int \phi_0^* \Omega^* f_j \mathcal{G}_\nu \, d^3r\Big)  \, \hat{a}_\nu \hat{A}_j.
\end{split}
\end{equation}
They have the same order of magnitude, $\sqrt{N_0Q_0}$.

The block $\epsilon_{ph}$ is just a diagonal $m_{ph}\times m_{ph}$ matrix filled with bare energies of high-Q mode photons, $\hbar \omega_{\nu}$.
The effective photon-photon interactions is described by the co-rotating photon-photon block $S_{ph}$ only, which is inherrited from the third summand of the Hamiltonian~(\ref{H-cavityQED}) and whose entries are proportional to the number of particles in the condensate $N_0$:
\begin{equation}
    (S_{ph})_{\nu\nu'} \, \hat{a}_\nu^\dagger \hat{a}_{\nu'} \equiv
    \frac{\hbar}{\Delta_a} N_0 \Big( \int |\phi_0|^2 \mathcal{G}_\nu^* \mathcal{G}_{\nu'} d^3r\Big) \, \hat{a}_\nu^\dagger \hat{a}_{\nu'}.
\end{equation}
Counter-rotating photon-photon terms are zero, as it is in the  Hamiltonian~(\ref{H-cavityQED}).

The atomic operators are standard for the Bogoliubov-type theory including both co-rotating blocks $\epsilon_{at}$, $S_{at}$ and counter-rotating block $\tilde{S}_{at}$, which are not diagonal, in general case, due to the condensate-mediated atom-atom scattering characterized by the interaction constant $g_a$:
\begin{multline}
    (\epsilon_{at})_{jj'} \hat{A}^\dagger_j \hat{A}_{j'} = 
      \bigg(\int f_j^* \Big(-\frac{\hbar^2}{2m} + V_{tr} -\mu
    \\    
      +2g_a \big(N_0|\phi_0|^2 + \la\nex\ra\big) f_{j'} \Big) d^3r \bigg)
      \hat{A}^\dagger_j \hat{A}_{j'},
\end{multline}
\begin{equation}
\begin{split}
    &(S_{at})_{jj'} \hat{A}^\dagger_j \hat{A}_{j'} = 
        \frac{\hbar}{\Delta_a}  \Big(\int f_j^* \, 
        |\Omega|^2 \, f_{j'} \, d^3r\Big) \hat{A}^\dagger_j \hat{A}_{j'};
     \\
    &(\tilde{S}_{at})_{jj'} \hat{A}_j \hat{A}_{j'} = \frac{g_a N_0}{2} \Big( \int f_j \,(\phi_0^*)^2 \, f_{j'} \, d^3r \Big) \hat{A}_j \hat{A}_{j'}.
\end{split}
\end{equation}
Here $\mu$ is a chemical potential of Bose atoms (introduced according to the Bogoliubov approximation), while $N_0|\phi_0|^2$ and $\la\nex\ra$ are mean density profiles of the condensate and the noncondensate, respectively.
Both co-rotating and counter-rotating terms generated by condensate-mediated atom interactions,  scale proportionally to the number of particles in the condensate $N_0$.
However, the coherent optical field acts like an additional external potential, and keeping that in mind we highlight a separate co-rotating block $S_{at}$ which magnitude is proportional to optical field intensity, that is to $Q_0 = \int |\Omega|^2 d^3r \big/ \int |\mathcal{G}_0|^2 d^3r$.

The steady state represented by Eqs.~(\ref{rho-eff}), (\ref{H-eff-2}) is a convenient anzats suitable for discussion of a new setup for the boson sampling process.
In general the steady state of a hybrid atomic-optical system doesn't have to be a Gaussian pseudo-thermalized state, such as the anzats~(\ref{H-eff-2}). 
It will be discussed in a separate study elsewhere.
However, Gaussian states are often treated as more “classical" then other quantum states \cite{Walschaers2021NonGauss} (for example, non-Gaussianity is considered as a resource for quantum computations, see~\cite{Genoni2010NonGauss,Takagi2018NonGauss,Shapiro2018NonGauss}). 
So, finding quantum advantage within a model of a Gaussian state would strengthens the case of the proposed hybrid atom-photon system: 
If the Gaussian model already predicts the statistics which is \#P-hard for computing, sampling from a more quantum state, actually existing in the system, should not be simpler in terms of computational complexity.

\customsec{The covariance matrix $G$ of the above hybrid system}
of excited atom states and high-Q optical modes involves blocks of normal ($\mathcal{N}$) and anomalous ($\mathcal{A}$) correlators,
\begin{equation}
\begin{split}  \label{G-def}
    G &\equiv
    \renewcommand{\arraystretch}{1.5}
    \left[
    \begin{array}{cc|cc}
            \mathcal{N}_{ph-ph} & \mathcal{N}_{ph-at} & 
                                \mathcal{A}_{ph-ph}^* & \mathcal{A}_{ph-at}^* \\
            \mathcal{N}_{ph-at}^\dagger & \mathcal{N}_{at-at} &
                                \mathcal{A}_{ph-at}^\dagger & \mathcal{A}_{at-at}^* \\\hline
            \mathcal{A}_{ph-ph} & \mathcal{A}_{ph-at} & 
                                \mathcal{N}_{ph-ph}^* & \mathcal{N}_{ph-at}^* \\
            \mathcal{A}_{ph-at}^\tp & \mathcal{A}_{at-at}  &
                                \mathcal{N}_{ph-at}^\tp & \mathcal{N}_{at-at}^* \\
    \end{array}\right]
    \renewcommand{\arraystretch}{1};
    \\
    &\mathcal{N}_{ph-ph} \equiv \big(\la \hat{a}^\dagger_\nu \hat{a}_{\nu'} \ra \big),
    \quad
    \mathcal{A}_{ph-ph} \equiv \big(\la \hat{a}_\nu \hat{a}_{\nu'} \ra \big),
    \\
    &\mathcal{N}_{at-at} \equiv \big(\la \hat{A}^\dagger_j \hat{A}_k \ra \big),
    \quad
    \mathcal{A}_{at-at} \equiv \big(\la \hat{A}_j \hat{A}_{k} \ra \big),
    \\
    &\mathcal{N}_{ph-at} \equiv \big(\la \hat{a}^\dagger_\nu \hat{A}_j \ra \big),
    \quad
    \mathcal{A}_{ph-at} \equiv \big(\la \hat{a}_\nu \hat{A}_j \ra \big).
\end{split}    
\end{equation}   
For a quasi-equilibrium state, 
$\hat{\rho} \propto \exp(-\hat{\mathcal{H}}_\eff/T_\eff)$, 
it can be expressed in terms of the Hamiltonian (\ref{H-eff-2}) as follows
\begin{equation}    \label{G-to-H}
    G = \frac{1}{2} \coth \bigg( \frac{ J H }{ 2T_\eff } \bigg) \, J - \frac{\mathbbm{1}}{2},
    \qquad
    J \equiv    \begin{bmatrix}
                    +\mathbb{1} & \mathbb{0} \\  
                    \mathbb{0}  & -\mathbb{1} 
                \end{bmatrix}.
\end{equation}
Here all blocks of the matrix $J$ are of a size $m_{at}+m_{ph}$.
A less explicit form of the relation~(\ref{G-to-H}), which is more transparent, especially in the limit $T_\eff \to 0$ corresponding to a multimode squeezed vacuum state, is as follows
\begin{equation}    \label{G-to-H-via-R}
\begin{split}
    G = &R    \begin{bmatrix}
                    \mathcal{N}^{(qp)} & \mathbb{0} \\  
                    \mathbb{0}  & \mathcal{N}^{(qp)}
                \end{bmatrix}
    R^\dagger + (R R^\dagger - \mathbb{1})/2,
    \\
    &\mathcal{N}^{(qp)} = \textrm{diag} \big(\{ \la n^{(qp)}_j \ra\}\big).
\end{split}    
\end{equation}
Here $R$ is a complex symplectic matrix representing the Bogoliubov transform linking the creation and annihilation atomic and photon operators to the ladder operators $\hat{B}^\dagger_j$ and $\hat{B}_j$ of quasiparticles, which are mixtures of atoms and photons that diagonalize the Hamiltonian~(\ref{H-eff-2})
\footnote{
The exact form of the Bogoliubov transform matrix $R$ may be determined from the Bogoliubov-de Gennes equation:
$$
    \begin{bmatrix}
        +\mathbb{1} &   \mathbb{0} \\
        \mathbb{0}  &   -\mathbb{1}
    \end{bmatrix}
    H R 
    =
    R \,
    \begin{bmatrix}
        +\textrm{diag}\big(\{E_j\}\big) &   \mathbb{0} \\
        \mathbb{0}  &   -\textrm{diag}\big(\{E_j\}\big)
    \end{bmatrix}.
$$
It follows from Eq.~(\ref{mH-diag}) in virtue of the identity 
$
    R^{-1} 
    = 
    \begin{bmatrix}
        +\mathbb{1} &   \mathbb{0} \\
        \mathbb{0}  &   -\mathbb{1}
    \end{bmatrix}
    R^\dagger
    \begin{bmatrix}
        +\mathbb{1} &   \mathbb{0} \\
        \mathbb{0}  &   -\mathbb{1}
    \end{bmatrix},
$
which holds since $R$ is a complex symplectic matrix of a special block structure, 
$R = \begin{bmatrix}
    U & V^* \\
    V & U^*
\end{bmatrix}$.
}
:
\begin{equation}    \label{mH-diag}
\begin{split}
    &({\bf a}^\dagger, {\bf A}^\dagger, {\bf a}, {\bf A} )^\tp = R \ ( {\bf B}^\dagger, {\bf B})^\tp,
    \\
    &R^\dagger \, H \, R =  
        \renewcommand{\arraystretch}{1.25}
                        \begin{bmatrix}
                                \textrm{diag} \big(\{ E_j \}\big) & \mathbb{0} \\  
                                \mathbb{0} & \textrm{diag} \big(\{ E_j \}\big)
                        \end{bmatrix}.
        \renewcommand{\arraystretch}{1}
\end{split}
\end{equation}
The bold symbols again denote the row-vectors, $\hat{\bf B} = (\hat{B}_1,\hat{B}_2\ldots)$; $E_j$ and $\la n^{(qp)}_j \ra \equiv \la \hat{B}_j^\dagger \hat{B}_j \ra = 1 \big/ \big(e^{ E_j/{T_\eff}} - 1\big)$ denote,respectively, the energies and mean occupation numbers of the quasiparticles which fluctuated independently since the other quasiparticles' pair correlators are zero.
Relatively large norms of the co- and counter-rotating intermode terms in the Hamiltonian lead to an efficient mixing of atomic and photon modes within the quasiparticles in virtue of the Bogoliubov transform $R$.

In order to analyze the multimode photon statistics only, disregarding the occupation numbers in the atomic modes, the covariance submatrix $G_{ph}$ characterizing only photon-photon correlators should be allocated within the whole covariance matrix $G$:
\begin{equation}
    G_{ph} =    
            \renewcommand{\arraystretch}{1.5}
                \begin{bmatrix}
                    \mathcal{N}_{ph-ph}     & \mathcal{A}_{ph-ph}^* \\
                    \mathcal{A}_{ph-ph}     & \mathcal{N}_{ph-ph}^* \\
                \end{bmatrix}
            \renewcommand{\arraystretch}{1}
    =
    \renewcommand{\arraystretch}{1.5}
        \begin{bmatrix}
                    \big( \la \hat{a}^\dagger_\nu \hat{a}_{\nu'} \ra \big)
                        & \big( \la \hat{a}^\dagger_{\nu} \hat{a}^\dagger_{\nu'} \ra \big)\\
                    \big(\la \hat{a}_\nu \hat{a}_{\nu'} \ra \big)   
                        & \big(\la \hat{a}^\dagger_{\nu'} \hat{a}_\nu \ra \big)\\
        \end{bmatrix}.
    \renewcommand{\arraystretch}{1}
\end{equation}
In the general case, the block of photon-photon anomalous correlators $\mathcal{A}_{ph-ph} = \big( \la \hat{a}_\nu \hat{a}_{\nu'} \ra \big)$ is nonzero for a Hamiltonian corresponding to the matrix $H$, Eq.~(\ref{H-eff-2}), although the Hamiltonian doesn't involve counter-rotating terms acting on photon modes only.
For the zero effective temperature, which is the most favorable regime for observing nontrivial quantum statistics, the anomalous photon correlators $\la \hat{a}_\nu \hat{a}_{\nu'} \ra$ appear due to the quantum-depletion summand $(R R^\dagger - 1)/2$ in the expression~(\ref{G-to-H-via-R}) for the covariance matrix.

According to the hafnian master theorem~\cite{LAA2022-HafnianMasterTheorem}, the photon covariance matrix $G_{ph}$ immediately provides joint probabilities of measuring a sample of photon numbers $\{n_\nu\}$ in terms of the matrix hafnian,
\begin{equation}    \label{P-ph-GBS}
    p(\{n_\nu\}) = \frac{\haf \tilde{C}(\{n_\nu\})}{\sqrt{\det (\mathbb{1}+G)} \, \prod_\nu n_\nu!}.
\end{equation}
Here a block matrix $\tilde{C} (\{n_\nu\})$ of dimension $(\sum_\nu n_\nu)\times(\sum_\nu n_\nu)$ under the hafnian is build from the  $2m_{ph}\times 2m_{ph}$ correlation-related matrix $C \equiv \begin{bmatrix}  \mathbb{0} & \mathbb{1} \\
                           \mathbb{1} & \mathbb{0} \end{bmatrix} G_{ph} (1+G_{ph})^{-1}$
according to the following prescription: 
Each entry $C_{\nu,\mu}$ at the position $(\nu,\mu)$, $1\le \nu,\mu \le m_{ph}$, is replaced by a $n_\nu\times n_\mu$ block filled with $C_{\nu,\mu}$;
the same is done for the entries in other three blocks of $C$, at the positions $(m_{ph}+\nu,\mu)$ as well as $(\nu,m_{ph}+\mu)$ and $(m_{ph}+\nu,m_{ph}+\mu)$.

It is also interesting to note, that the system naturally incorporates effective interferometers: According to the Bloch-Messiah reduction of the Bogoliubov-transform matrix
\cite{Braunstein2005-BlochMessiah,CariolaroPRA2016-BlochMessiahRotCond},
\begin{equation}
    R = \begin{bmatrix}
            U_1   &     \mathbb{0}  \\
            \mathbb{0}  &   U_1^*
        \end{bmatrix}\!\!
        \begin{bmatrix}
            \cosh r & \sinh r \\
            \sinh r & \cosh r
        \end{bmatrix}\!\!
        \begin{bmatrix}
            U_2   &     \mathbb{0}  \\
            \mathbb{0}  &   U_2^*
        \end{bmatrix}\!,
        \
        r = \textrm{diag}\big(\{r_j\}\big),
\end{equation}
it involves two nonequivalent (and nontrivial, in the general case) unitaries $U_1$, $U_2$. One of them affects statistics through the covariance matrix $G$ even in case of zero quasiparticle occupation numbers.
Both unitaries depend on various overlapping integrals, involving atomic wave functions and spatial profiles of the optical modes supported by the cavity, and thus could be controlled be changing geometrical properties of the system.

The probability distribution in Eq.~(\ref{P-ph-GBS}) means that the system should demonstrate Gaussian boson sampling.
For certain parameters of the system, which correspond to the considerable effect of quantum depletion and significant values of anomalous correlators, the involved hafnians are hard to approximate classically (due to exactly the same argumentation stated for pure optical setups; see, for example, \cite{Hamilton2019-DetailedGBS}). 

\customsec{A simple two-mode model,}
which proves that such sets of parameters, ensuring a dominant role of anomalous correlators, are definitely achievable, is presented below.


In order to illustrate the general concept described above we consider a simple toy model of a system involving just two modes, one “atomic” and one “photon”, interacting according to the quadratic Hamiltonian
\begin{widetext}
\begin{equation}    \label{H-toy}
    \hat{\mathcal{H}}_\eff = \frac{1}{2} \, \big(\hat{a}, \hat{A}, \hat{a}^\dagger, \hat{A}^\dagger \big)
        \begin{bmatrix}
            \hbar \omega + 2\gamma \sqrt{\tfrac{N_0}{Q_0}} & \gamma & 0 & \gamma \\
            \gamma & \epsilon + 2\gamma \sqrt{\tfrac{Q_0}{N_0}} & \gamma & 0 \\
            0 & \gamma & \hbar \omega + 2\gamma \sqrt{\tfrac{N_0}{Q_0}} & \gamma \\
            \gamma & 0 & \gamma & \epsilon + 2\gamma \sqrt{\tfrac{Q_0}{N_0}}
        \end{bmatrix}
        \renewcommand{\arraystretch}{1.55}
        \begin{pmatrix}
            \hat{a}^\dagger \\
            \hat{A}^\dagger \\
            \hat{a} \\
            \hat{A} 
        \end{pmatrix}
        \renewcommand{\arraystretch}{1}.
\end{equation}
\end{widetext}
Here the ladder operators $\hat{A}$ and $\hat{A}^\dagger$ refer to the only “atomic” mode with a bare energy $\epsilon$, and operators $\hat{a}$ and $\hat{a}^\dagger$ -- to the only “photon” mode with a bare energy $\hbar\omega$.
All matrix entries are taken to be real, which can always be achieved for such a matrix by means of a proper choice of the phases of atomic and optical modes (which cannot be done in the general case of multiple modes).

The pattern of the toy model Hamiltonian coincides with the effective quadratic Hamiltonian~(\ref{H-eff-2}) which describes the optically driven BEC system in a multimode cavity, Eq.~(\ref{H-cavityQED}).
Namely, the intermode co-rotating and counter-rotating terms are of the same magnitude, characterized by the parameter 
$\gamma \sim \frac{\hbar\la\mathcal{G}^2\ra}{\Delta_a}\sqrt{N_0Q_0}$,
while “photon-photon” counter-rotating terms are exactly zero.
The magnitude of corrections to the of diagonal entries is $N_0/Q_0$, where $N_0$ and $Q_0$ are, respectively, the occupations of the Bose condensate and the classical optical mode, on top of which the atomic and optical excitations described by the toy model (\ref{H-toy}) exist.
Moreover, the “atom-atom” counter-rotating terms are omitted, since in cavity-QED experiments atomic scattering on the condensate may be significantly weaker than the light-induced atomic interaction \cite{Ritsch2021-CavityQED}.
As is shown below, they are not crucial.


We assume that the system is in a Gaussian pseudo-equilibrium state, $\hat\rho \propto \exp( -\hat{\mathcal{H}}_\eff/T_\eff)$ at some effective temperature $T_\eff$ and consider the photon counting number statistics in the only optical mode.
Such statistics is fully determined by the “photon-photon” normal and anomalous correlators, denoted by $\eta \equiv \la\hat{a}^\dagger\hat{a}\ra$ and $\alpha \equiv \la\hat{a}\hat{a}\ra$, respectively.
For states close to the squeezed vacuum, such two-parameter statistics demonstrate nontrivial suppression of odd occupation number probabilities \cite{Weedbrook2012-GaussianReview, BarnettRadmore-QuantOpt, Entropy2023-ABSbox}. Being supplemented by the interference in the multimode case, such effects result into nontrivial, hard-to-mimic patterns in the Gaussian boson sampling.
For illustration purposes, we introduce a typical value of the anomalous correlator, $\alpha_c = \sqrt{\eta^2 +\eta/2}$, that should be exceeded to get significant squeezing effect on the photon statistics. 
This characteristic value of the anomalous correlator corresponds to approximately equal probabilities of one-photon and two-photon outcomes meaning that the odd occupation probabilities are visibly suppressed. 
At the same time, a fully squeezed quantum vacuum state is achieved when the anomalous correlator reaches its maximum possible absolute value $\alpha_{max} = \sqrt{\eta^2 +\eta}$.

\begin{figure}[p]         
    \centering
    \includegraphics[width=8cm]{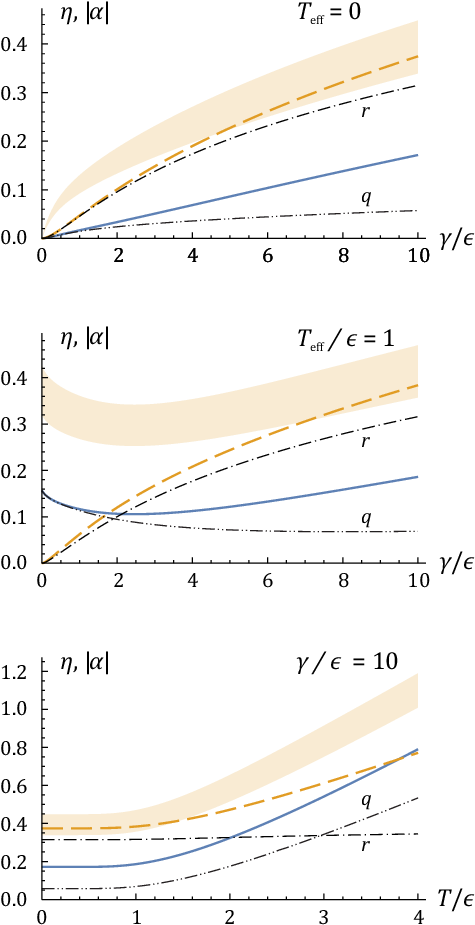}
    \caption{   \label{fig1:anom-vs-norm-corr}
    Dependence of the normal ($\eta$) and anomalous ($\alpha$) photon correlators in a generic two-mode model, Eq.~(\ref{H-toy}), on the interaction parameter $\gamma$, panels (a) and (b), as well as on the effective temperature of the system $T_\eff$, panel (c). 
    The normal correlator $\eta$ is given by the blue solid curves, while the absolute value of the anomalous correlator $\alpha$ is given by the yellow long-dashed curves.
    The shaded area marks the range of the anomalous correlator values $\big(\alpha_c,\alpha_{max} \big) = \big(\!\sqrt{\eta^2+\eta/2},  \sqrt{\eta^2+\eta}\,\big)$
    with a pronounced quantum effect of suppression of odd occupation probabilities.
    Dot-dashed curves show the effective single-mode squeezing parameter
    $r = \frac{1}{2}\textrm{artanh}\, \frac{|\alpha|}{\eta+1/2}$, and the effective mean occupation number before imposing squeezing, $q = \sqrt{(\eta+1/2)^2 - |\alpha|^2} - 1/2$.
    The main numerical parameters of the model are taken to be of the same order of magnitude: $\hbar\omega/\epsilon = 2$, $Q_0/N_0 = 7$.
    \pagebreak
    }     
\end{figure}

As is shown in Fig.~\ref{fig1:anom-vs-norm-corr}, the minimalistic toy model described above is sufficient for demonstrating efficient generation of the anomalous “photon-photon” correlator $\alpha$ by means of the counter-rotating atom-photon 
coupling~\footnote{
    In principle, the desired correlators (as well as the Bogoliubov transform which diagonalizes the Hamiltonian) and the whole correlation matrix may be calculated analytically, since the matrix dimension is only $4\times4$.
    However, the resulting formulae are cumbersome and not really transparent.
    Therefore, we omit these calculations and limit ourselves to numerical illustrations only.
}.
The absolute value of $\alpha$ crosses the aforementioned appearance level $\alpha_{c}$ -- and the single-mode photon statistics exhibits well-pronounced nontrivial features -- if the effective temperature is low enough, the light-atom interactions are significant, and the shift of the atomic energy in the diagonal term of the Hamiltonian is larger than that of the photon energy (that is, $Q_0 > N_0$).

The “photon-photon” anomalous correlator originates from two ingredients responsible for the relation between “photon” and “atomic” modes as well as quasiparticle and eigen-squeeze modes of the system -- the effect of two-mode squeezing (generated by the intermode counter-rotating terms in the Hamiltonian) and the effect of mode mixing (generated by the intermode co-rotating terms in the Hamiltonian) which is also crucial.
In terms of the general Hamiltonian~(\ref{H-cavityQED}), both these effects are governed by the same, second summand describing the interaction directly involving the coherent optical field.
In the absence of the intermode co-rotating terms, the squeezing effects are still present in the system, but they produce only intermode, “atom-photon” anomalous correlators, while “photon-photon” anomalous correlators remain zero, so that a single mode “photon” counting-number statistics is simply a thermal distribution. 
The latter immediately follows from the fact that the Bogoliubov transformation in that special case would have a very simple and restricted form $\hat{B}_{1} = \hat{A} \cosh r + \hat{a}^\dagger \sinh r$, $\hat{B}_{2} = \hat{a} \cosh r  + \hat{A}^\dagger \sinh r$, well-known for the two-mode squeezing \cite{BarnettRadmore-QuantOpt, VogelWeksch-Quantopt,Caves1982QuantLimits}.

Note that even when the whole two-mode system is in the vacuum state (i.e., $T_\eff\to 0$), the “photon” statistics doesn't coincide with that of the single-mode squeezed vacuum as is explained below.
Indeed, the analyzed “photon” mode sampling statistics can be fully reproduced by the statistics of a single mode squeezed to the amount $r = \frac{1}{2}\textrm{artanh}\, \frac{|\alpha|}{\eta+1/2}$ from the initial Gaussian state with an expected occupation number $q = \sqrt{(\eta+1/2)^2 - \alpha^2} - 1/2$. 
Hereinafter we refer to $r$ and $q$ as to the effective squeezing parameter and effective number of quasiparticles, respectively.
The single-mode squeezed vacuum statistics corresponds to the zero effective number of quasiparticles, $q=0$, which occurs when the $|\alpha|$ reaches the maximum possible value.
However, the value of the “photon” anomalous correlator in the considered atomic-photon system doesn't achieve it's upper boundary.
That means that tracing out the irrelevant atomic statistics not only retains the squeezing effect in the pure photon subsystem, but also results in a nonzero effective mean quasiparticle number $q$ in the squeezed single mode.
Even for $T_\eff = 0$ the anomalous correlator $\alpha$ at small values of the interaction $\gamma$ grows only quadratically in $\gamma$, as the normal correlator $\eta$ does, and remains much smaller than the boundary $\alpha_{max}$ which scales linearly in $\gamma$.

It is in contrast to the case of the single-mode squeezed vacuum, where the anomalous correlator in the same region grows linearly in $r$ -- and therefore in $\gamma$ -- and essentially overtakes the normal correlator which follows the quadratic law.
As a result, for a hybrid atom-photon system even at the zero effective temperature the nontrivial squeezing statistical effects reveal themselves only for the interaction strength achieving a certain level.

At the same time, the effect of generating the “photon-photon” anomalous correlator is a rather robust effect.
It does not rely on some specially selected values of the system parameters, such as the ratio of $\hbar\omega/\epsilon$, and a variation of the parameters in a quite wide ranges doesn't lead to qualitative changes regarding the existence of the suppression-of-odd-occupations effect.
Increasing $T_\eff$ from the zero value also leads only to smooth changes in the region of significant interactions. A sharp restructuring of correlators' behavior occurs only in a region of weak interactions, where the anomalous correlator is not strong enough anyway.
The main parameter that determines to which extent the anomalous correlator $|\alpha|$ can approach its upper boundary, $\alpha_{max} = \sqrt{\eta^2+\eta}$,
is the ratio $\sqrt{N_0/Q_0}$, and the value of $Q_0$ should somewhat prevail over $N_0$.
If the ratio $N_0/Q_0$ is about $1$ or smaller, a sufficiently close rapprochement never happens at any magnitude of the interaction parameter $\gamma$ since an effective number of quasiparticles $q$
grows too fast with enlarging the interaction.  
Nevertheless, even if the backward inequality $Q_0 < N_0$ holds, a proper increase of the anomalous “photon-photon” correlator values is still achievable --- in particular, by a simple modification of the toy model via adding several more “atomic” modes (which still interact with the single “photon” mode only).

In all of the above discussions we neglected by the “atom-atom” counter-rotating terms in the Hamiltonian (which are present in a weakly interacting gas, but may be not of the leading order of magnitude).
Obviously, introducing these terms further enhances the squeezing effect in the system, and thus only supports generating anomalous correlators.
Hence, the picture outlined above qualitatively persists.
Importantly, as we have seen above, the presence of these “atom-atom” counter-rotating terms is not necessary for generation of the anomalous “photon-photon” correlators required for appearance of the nontrivial photon number statistics.

\customsec{Concluding remarks.}
We propose to leverage the multimode squeezing naturally occurring in hybrid atom-photon multimode systems as a mean to generate \#P-hard joint counting photon statistics in the boson sampling experiments aimed at manifestations of quantum advantage.
This is in contrast to a conventional approach, in which input squeezed or Fock states are generated by some external sources and then should be properly synchronized.
In the suggested concept, both the generation and synchronization are established naturally due to the atom-photon interaction in the system, which is crutially important for a scalability of boson sampling experiments.

The proposed idea is outlined for a hybrid atom-photon system of partially-condensed Bose gas trapped inside a multimode cavity. 
We show that if such a system contains, along with the Bose condensate, a macroscopically populated optical coherent mode, than the pseudo-equilibrium multimode statistics of a photonic subsystem is characterized by the strong anomalous correlators.
That happens even in the absence of counter-rotating terms in the Hamiltonian of the photon subsystem (which is the case for the ultracold gas placed in the multimode optical cavity). 
The combination of the counter- and co-rotating terms representing interaction of subsystems is enough to establish the effect, and counter-rotating terms acting on the atomic subsystem aren't necessary. 
We also show, via a simple two-mode model, that these anomalous correlators may be large enough to enable quantum squeezing effects in the photon number statistics that result, in the multimode case, in the hard-to-mimic joint probability patterns of a Gaussian boson sampling.

The obtained results aren't restricted to measuring photon numbers.
Instead of considering joint statistics of a photon subsystem only, one may study a mixed counting number statistics of any set of both atomic and optical modes, which have in general the same properties.
We have mostly focused on photons only because measurements of optical states are well developed and more easily accessible in experiments.
Within already existing quantum-gas and cavity-QED technologies~\cite{cavityQED-Humb,cavityQED-Stanford,cavityQED-ETH1,cavityQED-ETH2}, the proof-of-principal experiments of demonstrating hafnian-featured statistics of photons in a few mode regime look feasible.
\\

{\it We acknowledge the support by the Center of Excellence “Center of Photonics” funded by The Ministry of Science and Higher Education of the Russian Federation, contract № 075-15-2022-316.}

\bibliographystyle{unsrt}
\bibliography{list}

\end{document}